\documentclass[a4paper,aps,manuscript]{revtex}
\tightenlines
\usepackage[T1]{fontenc}
\usepackage[dvips]{graphics}

\makeatletter

\newcommand{\LyX}{L\kern-.1667em\lower.25em\hbox{Y}\kern-.125emX\spacefactor1000}

\makeatother

\begin{document}

\title{On the Rapid Estimation of Permeability for Porous Media Using Brownian Motion Paths}

\author{Chi-Ok Hwang}

\address{Department of Computer Science, Florida State University, 203 Love Building Tallahassee, FL 32306-4530}

\author{James A. Given}

\address{Angle Inc., 7406 Alban Station Court, Suite A112, Springfield, VA 22150 }

\author{Michael Mascagni}

\address{Department of Computer Science, Florida State University, 203 Love Building Tallahassee, FL 32306-4530}

\maketitle
\begin{abstract}
We describe two efficient methods of estimating the fluid
permeability of standard models of porous media by using the statistics of continuous Brownian motion paths that initiate outside a sample and terminate on contacting the porous sample. 
The first method associates the "penetration depth" with a specific property of the Brownian paths, 
then uses the standard relation between penetration depth and permeability to calculate the latter. 
The second method uses Brownian paths to calculate an effective capacitance for the sample, 
then relates the capacitance, via angle-averaging theorems to the translational hydrodynamic friction of the sample. 
Finally, a result of Felderhof is used to relate the latter quantity to the permeability of the sample. 
We find that the penetration depth method is highly accurate in predicting permeability of porous material. 
\end{abstract}

\newpage
\section{Introduction}

Much theoretical effort has been expended in attempts to either estimate or bound the fluid 
permeability of a porous medium, given its average statistical properties. 
Techniques used for this include the volume-to-surface-area ratio \cite{vs}, the \(
  \Lambda  \) parameter \cite{lamda}, percolation ideas \cite{percolation}, and concepts
  from the theory of Brownian motion such as mean survival distance or mean survival time 
(inverse reaction rate)~\cite{Scheidegger,mst,Avellaneda}. 

  One basic class of theoretical models allows one to study either packed beds or consolidated porous media such as sandstone. These models consist of ensembles of equal-sized impermeable spherical inclusions immersed in a completely permeable medium. When used to model a packed bed, 
these spheres are non-overlapping; when used to model a consolidated porous medium, they are randomly located and freely overlapping. A second basic class of models for packed beds and/or porous media consist of periodic arrays of impermeable spherical inclusions
immersed in a freely permeable medium. Depending on the ratio between the diameter of the 
spherical inclusions and the separation between neighboring inclusions, and the periodic lattice chosen, these models may consist either of overlapping or non-overlapping spheres. 

The use of Monte Carlo methods to solve elliptic or parabolic partial differential equations (PDEs) is grounded in the mathematical subject called probabilistic potential theory\cite{Friedlin,Chung,Sznitman}. It is well known that the probability density function of a diffusing particle obeys the diffusion equation. The time-integrated, or steady-state probability density function obeys the Laplace equation. These facts have been greatly generalized. One can solve a large class of elliptic PDE's by simulating the corresponding class of (in general) biased diffusion processes~\cite{Friedlin,Sabelfeld}. The diffusion Monte Carlo method is 
defined by the mapping of boundary value problems involving elliptic PDE's into the corresponding diffusion problems, and
solving them by using Monte Carlo methods. This method is
powerful because it can be implemented computationally in a
massively parallel manner.

  Previous work\cite{Given} involving one of us (J.G.) develops a class of diffusion Monte Carlo algorithms, the Green's function based first-passage (GFFP) methods that are very efficient for studying the properties of random or highly irregular two-phase media. This class of problems involves:

\noindent$\bullet$ Interfacial surfaces that may pass very close to each other, thus creating channels that are narrow, but nevertheless important for understanding the bulk properties of the material.

\noindent$\bullet$ Spatial intermittency: large regions with no interfacial surface; smaller regions with a high density of interfacial surface. Intermittency may be due e.g. to polydispersity in pore diameter or void size, or to a dispersion, in solution, of very large irregular inclusions, such as macromolecules.

\noindent$\bullet$ Interfacial surfaces that may include sharp corners and edges. The difficulties these features cause in solving the Laplace equation are clear (In electrostatic terminology, 
they cause singularities in charge density.). 

\noindent In particular, the class of GFFP algorithms,\cite{Given}:

\noindent$\bullet$ Uses a purely continuum description of both sample and diffusing particle trajectory, with no discretization either in space or in time and hence no discretization error. This is essential for calculations of permeability in low-porosity materials.

\noindent$\bullet$ Uses a library of Laplace Green's functions that allow a diffusing particle both to make large `jumps' across regions containing no interfacial surface,
and to be absorbed from a distance, at an absorbing boundary, in a single jump.

\noindent$\bullet$ Uses `charge-based' rather than `voltage-based'
methods of calculation, i.e., retain only the location of absorption events, or equivalently surface charges. This allows one to perform trajectory simulation using importance sampling (because each trajectory yields one surface charge); thus ensuring that even singularities associated with corners and edges will be efficiently sampled. 

\noindent 
This method is an adaptation and refinement of the "walk on spheres" (WOS) method of diffusion Monte Carlo~\cite{Sabelfeld}. This method has been used since at least 1956 to treat diffusion problems~\cite{Muller}. It was first applied in random-media problems by Zheng and Chiew~\cite{Zheng}; see also~\cite{Torquato-Kim}. Other recent studies have also generalized WOS to diffusion in two-phase media modeled as arrays of small homogeneous squares~\cite{Siegel} and cubes~\cite{TKC};these studies use the simplest nontrivial Green's functions to determine first-passage probability. The Einstein relation may be used to obtain effective electrical conductivities~\cite{Siegel,TKC,SB} from the bulk diffusion coefficient of these models.

  The present method is more efficient than WOS for diffusing
particles very close to absorbing boundaries; as we show here, this makes it ideal for the study of low-porosity materials. 
It has been applied to solve the scalar Laplace equation, which governs both electrostatic and diffusion problems. 

  Electrostatic applications involve the calculation of electrostatic capacitance for very general bodies~\cite{BD,Douglas,Zhou}. This is an 
important set of problems because probabilistic potential theory implies fundamental relations between the capacitance and the lowest eigenvalue of the Laplacian in the Dirichlet problem~\cite{Douglas}. This in turn leads to efficient bounds or approximations for a wide variety of physical quantities in many-body systems, for example, the quantum mechanical scattering length in nuclear matter, or the classical rate of heat release in composite media. 

  Diffusive applications of this class of methods involve solutions of the Smoluchowski equation, either in the vicinity of a single irregular object, (e.g., a ligand diffusing near a reactive site on a macromolecule)~\cite{BD}, or in a system containing many absorbing objects, (e.g., systems with distributed trapping)~\cite{Zheng}. 

  Because so many physical phenomena are governed by vector Laplace equations (e.g. low-velocity fluid dynamics)~\cite{HB}, and tensor Laplace equations (e.g. solid mechanics)~\cite{Sabelfeld2}, it is natural to seek methods for extending this previous work to deal with these two classes of problems~\cite{Sabelfeld}.  An approach to  the first class is provided by the method of angle-averaging \cite{HD}. This method provides an approximate relation between the translational hydrodynamic friction \( f\) of an arbitrary body, and the electrostatic capacitance \(C\) of a perfect conductor of identical size and shape. For arbitrary nonskew objects with a no-slip boundary condition, using angular averaging of the Oseen tensor\cite{Wiegel}, Hubbard and Douglas \cite{HD} generalized 
the exact Stokes-Einstein result for the translational hydrodynamic
friction of a sphere,

\begin{equation}
 f_{sphere}=6\pi \eta R,  \label{eq1} 
\end{equation}

where \( \eta \) is the solvent viscosity and \(R\) is the radius of the sphere, to the translational hydrodynamic friction of an arbitrary impermeable body to give:

\begin{equation}
 f=6\pi \eta C  \label{eq2}
\end{equation}
where \(C\) is the corresponding capacitance.
This formula is exact in the "free tumbling" approximation, i.e., in the case that there is no coupling between the translational and rotational motion, and
that there is sufficient noise, or disorder, in the flow to cause
the object to occupy all possible orientations equally. But
extensive tests show the method is accurate well beyond this set of
cases.   

  A central insight of the work described here is this: the angle-averaging method may be applied to approximate the translational friction coefficient of an "object", even if the object is not a connected set, e.g., if it is a collection of inclusions that constitute the impermeable or matrix phase of a sample of porous media. A group of well-studied models for porous media fit this description. The inclusions may either be randomly placed, or set in a periodic lattice. The former models have properties similar to many commercially important porous media; the latter model a set of media formed by grain consolidation. The inclusions may be freely overlapping or impenetrable, i.e., unable to overlap. Finally the (spherical) inclusions may be uniform (monodisperse) in radius, or they may be polydisperse.  

  In this paper, we present two classes of accurate, computationally efficient methods of calculating permeabilities for these models
and models like them. These models combine rapid efficient methods of simulating 
Brownian motion \cite{Given,BD,Zhou}
with a pair of methods for deriving the permeability from the
statistics of Brownian particles diffusing near a sample of the
porous medium. The trajectories of these particles initiate outside the sample and terminate on contacting the porous matrix.
The spherical collection of small inclusion spheres will occasionally be referred to as the porous sample
in subsequent discussions.
The first method, the "penetration depth" method, associates the fluid-dynamic penetration depth with a specific property of the Brownian paths, then uses the standard relation between penetration depth and permeability to calculate the latter. The second method, the "unit capacitance" method, involves using Brownian paths to calculate an effective capacitance for the sample, then relates the capacitance, via angle-averaging theorems, to the translational hydrodynamic friction of the sample. Finally, a result of Felderhof~\cite{Felderhof} is used to relate the latter quantity to the permeability of the sample. 

  This paper is organized as follows: Section II sketches the method used here to obtain the electrostatic capacitance using Brownian paths. Section III develops the penetration depth method of calculating permeability, in which we obtain the penetration depth from Brownian paths, 
and obtain the permeability from the penetration depth. Section IV 
describes the unit capacitance method of calculating permeabilities, and applies it to the random models of 
packed beds and consolidated porous media. Section V applies both methods to determining the permeabilities
of the random models of packed beds. 
Section VI applies both methods to determining the permeabilities
of model packed beds composed of periodic arrays of impermeable spheres. 
Section VII applies both methods to determining the permeabilities of polydisperse randomly packed bed.
Section VIII contains our conclusions and suggestions for further study.

\section{Calculation of the Capacitance of a Sample of General Shape Using Brownian Paths}

  In this section, we sketch the algorithm for calculating the electrostatic capacitance of 
an irregularly shaped conducting body by using Brownian trajectories that initiate 
on a spherical launch surface surrounding the conducting body, and terminate on contact with that body. 
We focus this derivation in several ways in order to apply it to samples of porous media. 
If a packed bed model is either randomly packed or densely packed, any smooth sample boundary will intersect 
some of the spherical inclusions that constitute the sample; we provide two methods for dealing with this problem in section IV, the "effective radius method" and the "sharp boundary method". Difficulties will occur if the sample is chosen either too large or too small; we give an analysis 
that yields the acceptable qualitive range of sample sizes in section IV. For periodic models
of porous media, a non-round, e.g. cubic, sample (and thus a non-round launch surface) might be more appropriate. 
The derivation provided here is readily generalized to allow a cubic launch surface; in forthcoming work we explore this possibility numerically.

The electrostatic potential \(\phi(r)\), at position \(r\), in the presence of a conducting object, and the probability density \(p(r)\) 
in the associated Smoluchowski diffusion problem, i.e., the probability
associated with finding a diffusing particle at position \(r\), in the presence of an absorbing object of size and shape identical to the conductor, are related by:
\( \phi(r) = 1 - p(r)\). 
\noindent
This equation is derived by noting that its LHS and RHS obey the Laplace equation with identical boundary conditions. 

Thus the capacitance $C$ of a conducting object is given by:

\begin{equation}
C = -(4\pi)^{-1} \oint_{\Omega} d\sigma\cdot\nabla\phi(r) = (4\pi)^{-1}\int_{\Omega}d\sigma\cdot\nabla p(r).
  \label{Capa}
\end{equation}
where \(\Omega\) is the surface of the object.

The diffusion-controlled reaction rate $\kappa$ is defined as~\cite{BD,Zhou1,Zhou2}

\begin{equation}
\kappa = D \oint d\sigma\cdot\nabla p(r)    \label{DCR}
\end{equation}

\noindent
where the integral is evaluated on any closed surface containing the inner boundary \( \Omega \) 
(because of Green's theorem any such surface may be used.).

The relation between the diffusion-controlled reaction rate, in the Smoluchowski problem, 
of particles absorbed upon contact with an object, and the electrical capacitance of a conductor 
with identical size and shape to that object is thus given by:

\begin{equation}
\kappa = 4 \pi DC   \label{generalized}
\end{equation}

\noindent

We define \( \beta_{\infty}(\vec{r};\Omega) \) to be the probability that the diffusing particle started at
\( \vec{r} \) will be absorbed on \( \Omega \) and 
\( \gamma_{\infty}(\vec{r};\Omega) \) the probability that the diffusing particle started at \( \vec{r} \) will go to infinity, i.e., 
that it will not be absorbed in finite time. Thus:

\begin{equation}
\gamma_{\infty}(\vec{r};\Omega) = 1 - \beta_{\infty}(\vec{r};\Omega).    \label{Gamma}
\end{equation}

\noindent
\( \gamma_{\infty}(\vec{r};\Omega) \)
can be shown to satisfy the Laplace equation~\cite{Zhou2}. If the obvious boundary conditions,
\( \gamma_{\infty}(\infty;\Omega)= 1 \) and 
\( \gamma_{\infty}(\vec{r};\Omega)= 0 \) for \( \vec{r} \) on
\(\Omega \) are considered, it is clear that

\begin{equation}
p(r) = \gamma_{\infty}(\vec{r};\Omega).      \label{pofr}
\end{equation}

Using Eqs.~\ref{Gamma} and \ref{pofr}, Eq.~\ref{DCR} becomes:

\begin{equation}
\kappa =-D\oint_{r} d\sigma\cdot\nabla \beta_{\infty}(\vec{r};\Omega)
\end{equation}

\begin{equation}
=-D\oint r^{2} \sin \theta d\theta d\phi \frac{\partial}{\partial r}\beta_{\infty}(r;\Omega),
\end{equation}

\begin{equation}
=-r^{2}D \frac{\partial}{\partial r} \oint \sin \theta d\theta d\phi \beta_{\infty}(r;\Omega), \label{Constant}
\end{equation}

Defining

\begin{equation}
\beta(r;\Omega) = \frac{1}{4\pi} \oint \sin \theta d\theta d\phi \beta_{\infty}(r;\Omega),
\end{equation}

\noindent
we obtain from Eq.~\ref{Constant}

\begin{equation}
-4 \pi r^{2}D \frac{\partial}{\partial r} \beta(r;\Omega)= \kappa. \label{Integral}
\end{equation}

For any spherical surface \( r = b \) which contains the boundary \(\Omega\), integrating Eq.~\ref{Integral} from \( b\) to \( r\) with respect to \( r \) and using the boundary condition \( \beta( \infty ; \Omega) = 0 \), gives an expression for the reaction rate \( \kappa \):

\begin{equation}
 \kappa = 4\pi Db\beta(b;\Omega)   \label{above}
\end{equation}

Comparing Eqs.~\ref{generalized} and ~\ref{above} yields an important formula for
the electrostatic capacitance 
\begin{equation}
C=b\beta(b;\Omega)
\end{equation}

\begin{equation}
\equiv b\beta  \label{Beta}
\end{equation}

where \(\beta\) is the fraction of diffusing particles started at a random, i.e., angle-averaged, position on the launch sphere that are absorbed on the target. The unit
capacitance method, as described here, uses this relation to determine the capacitance
of a packed bed by performing Monte Carlo simulations for the quantity \(\beta\). Note that no assumption is made in the above that the "object" being studied is a connected set; it may be taken to be the set of inclusions that constitute the matrix phase of a sample of porous media.

We calculate the quantity $\beta$ by performing simulations as follows: each random diffusion path is constructed as a series of first-passage propagation jumps, each from the present position of the particle to a new position on a first-passage surface drawn around the present position. The new position is sampled from a first-passage position distribution function $f(x)$ which gives the probability density associated with finding a diffusing particle leaving the present position and first contacting the first-passage surface at point $x$. 

  This simplest first-passage surface is the first-passage sphere, i.e., a large sphere centered around the present position of the diffusing particle, that does not intersect any of the inclusions. The first-passage position density on this surface is uniform, i.e., its distribution function is trivial. Using only this first-passage surface yields a trivial case of the GFFP method, namely the WOS method. In this method, whenever a diffusing particle gets close enough (within a fixed distance $\delta_H$) to one of the inclusions,
it is taken to be absorbed. 

  We show here that for the problems under study it is more efficient to use more complex
first-passage surfaces. We let the first-passage sphere intersect the nearest inclusion,
and grow as large as possible, provided that it:

\noindent$\bullet$ not intersect the next-nearest inclusion

\noindent$\bullet$ not intersect any corners or edges of the nearest inclusion

The resulting first-passage surface is the portion of the first-passage sphere not
contained in any inclusion. Its surface consists of part of the first-passage sphere, and 
part of the surface of the inclusion. The probability distribution function $f(x,\vec P)$, for a diffusing particle starting at the center of the first-passage sphere and making first-passage at the point $x$ is in general quite nontrivial. (Here $\vec P$ is the set of geometric parameters that characterize the first-passage surface. Also, we assume for simplicity, that symmetry allows specification of the first-passage position $x$ by a single position parameter, the polar angle $\theta$.) Thus, $(\vec P, \theta)$ is the full set of parameters that the first-passage distribution function depends on. 

For a wide class of inclusions, the resulting first-passage surfaces are
'locally simple' closed geometrical surfaces. These surfaces are those for which we can tabulate, as a function of several parameters, the corresponding first-passage distribution function. To obtain this function, we: 

\noindent$\bullet$ Calculate the gradient, normal to the absorbing surface, of the appropriate Laplace Green's function for a grid of parameter and field values. 

\noindent$\bullet$ Calculate the indefinite integral of this gradient as a function of
$\cos\theta$. Normalize the integral to be unity:

\begin{equation}
I(\vec P, \cos\theta) = 1
\end{equation}

for $\cos\theta=1$. 

\noindent$\bullet$ Invert this relation for each set of parameter values $\vec P$ 
to obtain $\cos\theta$ as a function of $\vec P$ and $I$. The first-passage position, i.e., $\cos\theta$, can then be importance-sampled by choosing a random number $\alpha\in[0,1]$,
setting $I=\alpha$, and interpolating this relation.

If the chosen first-passage position is on the portion of inclusion surface included in the first-passage surface, the particle is absorbed. Otherwise, it will
reach a point on the rest of the first-passage surface. From there it makes its next jump. 
The diffusing particle continues until it is absorbed in the absorbing target or goes to infinity.

\begin{figure}[b]
\vspace{0.3cm}
{\centering \includegraphics{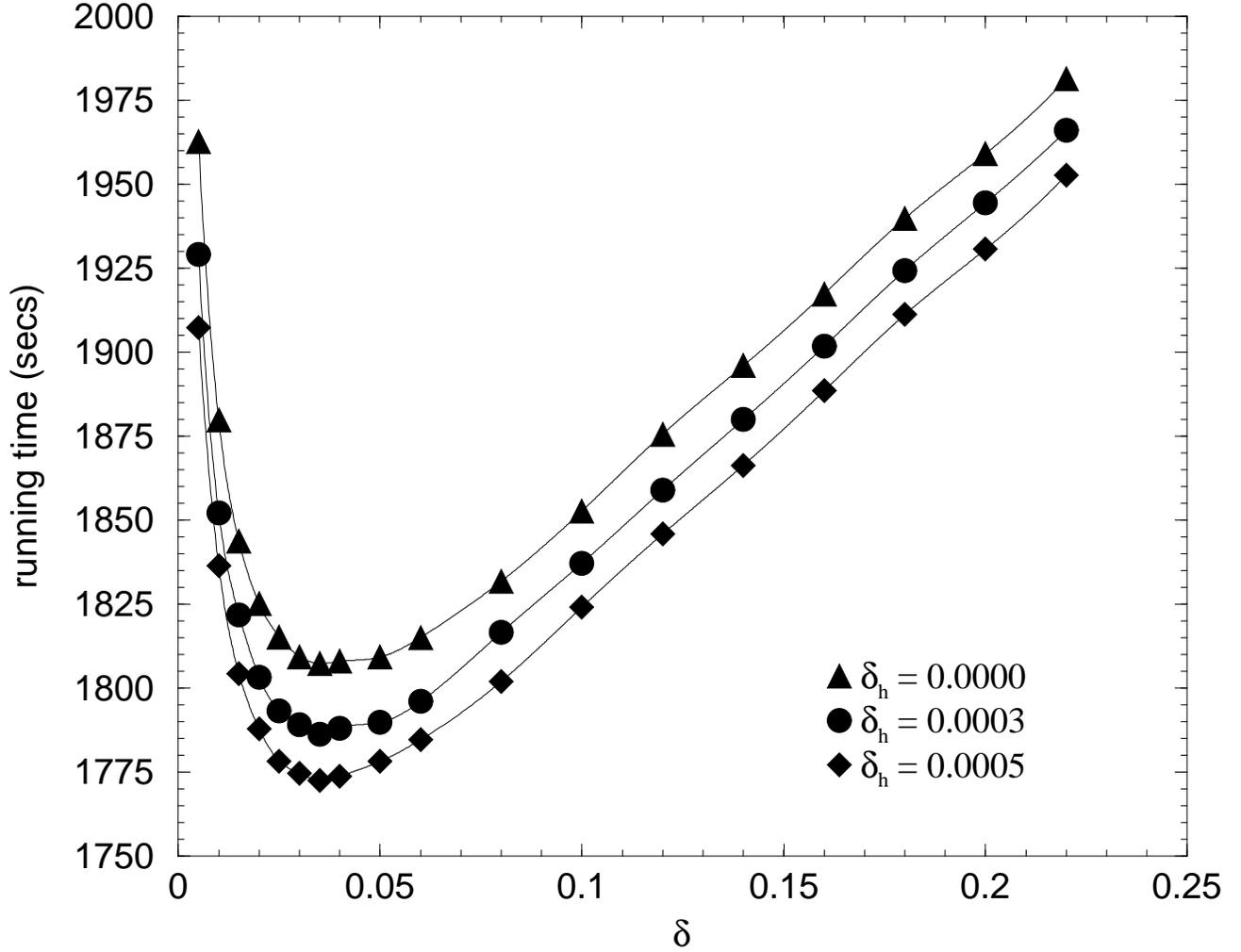} \par}
\vspace{0.3cm}
\caption{CPU time required to calculate the capacitance of the
unit cube, to a fixed tolerance, using the first-passage method.
The variables \(\delta_H\) and \(\delta_I\) prescribe the
distance from the surface of the cube that a diffusing particle must be, resp., to declare that a "hit", i.e., an absorption event has occurred, or to use a first-passage surface that intersects the surface of the cube. The figure shows that, for each value of \(\delta_H\), there is a non-zero optimal value of \(\delta_I\).
The walk-on-spheres (WOS) method is given by \(\delta_I=0\).}
\label{Runtime}
\end{figure}

As shown in Fig.~\ref{Runtime}, the GFFP method is more efficient than the 
WOS method whenever a small distance \(\delta_H\) must be used; this is true because the average number of moves required for absorption in the WOS method increases rapidly as  
\(\delta_H\) decreases \cite{Sabelfeld}. 
Thus, it is more efficient for studying porous
media either at low porosities or high polydispersity.

The set of tabulated first-passage distribution functions is built up by a process of `bootstrap diffusion Monte Carlo', in which simple distribution functions are used to accelerate the simulations used to calculate more complex distribution functions, and so on. In practice, the set of distribution functions is limited mainly by our inability to store in cache memory a tabulated distribution function depending on more than three parameters. 
In the present study the inclusions are asymmetric lenses (in the sharp-boundary method)
and spheres. These are readily treated by the GFFP method~\cite{Given,MacDonald}.

\section{The Penetration Depth Method of Calculating Permeability from the Absorption Point Statistics of Brownian Trajectories}

  In this section, we describe the `penetration depth' (PD) method for calculating permeability of an absorbing sample. First we relate the fluid-dynamic concept of "penetration depth" to a property of the paths of Brownian particles that diffuse from outside the sample of
porous media and terminate on contact with it. Then we use the standard relation between the permeability of a sample and its penetration depth to calculate the latter. 

  A simple model calculation allows one to relate the permeability of a porous medium to the effective penetration depth 
of fluid into the medium\cite{Wiegel}. Consider a homogeneous porous medium in a half-space 
with constant permeability \( k \). Fluid permeates this medium and has a constant velocity \( V_{0} \) within it. If \(z\) is the distance between the macroscopic fluid-medium interface and a point of the porous medium, the Debijf-Brinkman equation\cite{Brinkman} 

\begin{equation}
\nabla P=\eta\nabla^{2}{\bf V}-\frac{\eta}{k}{\bf V}. \label{Debijf}
\end{equation}

where \({\bf V}\) and \( P \) are macroscopic velocity and pressure respectively,

\noindent
reduces to the following,

\begin{equation}
\eta\frac{d^2V}{dz^2}= \frac{\eta}{k}V,
\end{equation}

\noindent
with the solution

\begin{equation}
V(z) = V_{0} \exp(-z/\sqrt{k})
\end{equation}
 
\noindent

This result shows that \( \sqrt{k} \) measures the distance that the flow effectively penetrates into the porous medium. We will call \( \sqrt{k} \) the `penetration depth'.

We can use this result to measure the permeability of a sample by identifying the penetration depth with the difference \( l \), between the average radial position at which the diffusing particles are absorbed
and the actual sample radius, thus yielding the approximate relation:

\begin{equation}
k = l^2
\end{equation}

The penetration depth defined here for diffusing particles is different from the mean linear survival distance: 
the latter is the average distance from the random starting point in the void phase to the absorption point; 
the former is the average of the radial component of this distance. It is possible that the mean linear survival distance 
can give an upper bound on permeability analogous to the Torquato-Kim upper bound~\cite{Torquato-Kim} based on the mean survival time. 

\section{The Unit Capacitance Method of Calculating Permeability of Random Systems}

  In this section, we describe an algorithm, the unit capacitance (UC) method, for calculating the permeability of a sample of random media. 
We calculate the capacitance of a sample (as described in Section II), use Eq. 
\ref{eq2},  from angle-averaging to give the translational hydrodynamic friction in terms of the capacitance, and finally use a relation, first published by Felderhof, between the translational hydrodynamic friction and the permeability to determine the latter.  

  Assuming that a given spherical sample is much larger than either the average distance between spherical inclusions or the correlation length associated with the statistics of the packed bed, it can be modeled as a homogeneous porous sphere with the appropriate porosity. 
One can solve the linear Stokes equation \cite{Wiegel,Felderhof} for the translational frictional coefficient, \(f\), of such a sphere by making the assumption that the excess pressure is linear in the fluid velocity, 
and by using the symmetry of the problem. The result is: 
\begin{equation}
 f=6\pi \eta RG_{0}(\sigma )\{1+\frac{3}{2\sigma ^{2}}G_{0}(\sigma )\}^{-1}  \label{eq3}
\end{equation}

where $\eta$ is the fluid viscosity and the function \(G_0(\sigma)\) is given by:

\begin{equation}
 G_{0}(\sigma )=1-\frac{1}{\sigma }\tanh \sigma . \label{eq4}
\end{equation}

Here \( \sigma  \) is the dimensionless quantity defined by

\begin{equation}
 \sigma =\frac{R}{\sqrt{k}},\label{eq5} 
\end{equation} 

where \(R\) is the porous sphere radius and \(k\) is the permeability.

Eliminating the translational frictional coefficient between Eqs.~\ref{eq2} and~\ref{eq3}, one finds a relation between the capacitance and the permeability

\begin{equation}
 \frac{C}{R}=G_{0}(\sigma )\{1+\frac{3}{2\sigma ^{2}}G_{0}(\sigma )\}^{-1}.\label{eq6}  
\end{equation}

Obtaining \(C/R\) (unit capacitance) from simulation allows  us to use Eq.~\ref{eq6} to calculate \(\sigma\), and thus obtain the desired
permeability estimate from Eq. ~\ref{eq5}.

\begin{figure}
\vspace{0.3cm}
{\centering \includegraphics{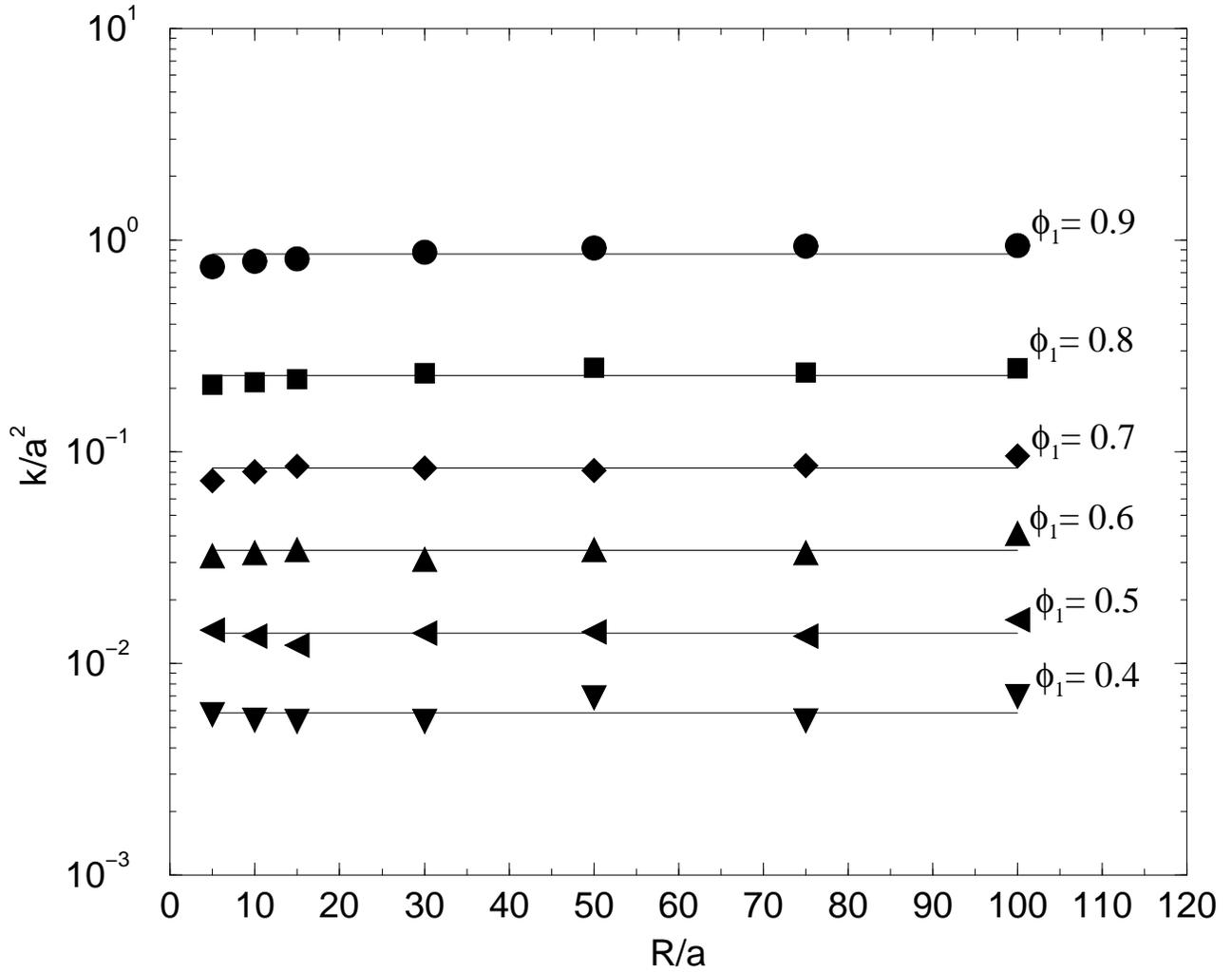} \par}
\vspace{0.3cm}
\caption{ Permeability vs \protect\(R/a\protect\) for unit capacitance estimation methods applied
to the randomly overlapping
sphere model: the permeability estimate does not depend on sample radius even though with small sampling
sizes more porous samples are needed. For \protect\(R=5, 10\protect\), 50 porous samples were used.When \(\phi_1 \) decreases, estimation is not stable due to the estimation instability. The solid lines are
the average of the seven permeabilities from different sampling sizes.}
\label{Flat1}
\end{figure}

  An important finding of the present work is that, even though neither the graph of C/R vs. R, 
nor that of  \(\sigma\) vs. C/R contains a flat region, the composition of the two, giving our permeability estimate as a function of R, does contain a flat "plateau" region.
For example, Fig.~\ref{Flat1} shows that for a broad range of porosities there is a plateau range of sample radii over which the permeability estimate does not vary. The lower
and upper limits of this region vary systematically.

In monodisperse homogeneous porous media models, the permeability is dependent 
only on the porosity (defined as the ratio 
of void space volume to total volume) and the inclusion sphere radius. 
By dimensional analysis, it can be expressed in the following functional form:

\begin{equation}
k = f(\phi_{1})a^{2}
\end{equation}

where \( \phi_{1} \) is porosity and \( a \) is inclusion sphere radius. 
The permeability estimates produced by the UC method will, in addition, depend on the parameter $R/a$. However, once this parameter is set larger than both the correlation length of the medium (very small in random media) and the average distance between inclusions, this
dependency should become quite weak. 

In Eq.~\ref{eq6}, for a given porosity and inclusion sphere radius (i.e. the permeability is constant), if we increase the sample radius, the unit capacitance goes to unity as \( \sigma \) goes to infinity. For each porosity, there is a range of sample radii which are much bigger than the average distance between spherical inclusions, but for which the ratio \(C/R\) will be far enough from unity to permit us to interpolate the corresponding \(\sigma\) value. We choose a radius that is in this range. 

When C/R is close to unity, that is, for very low porosities, the \(\sigma\) interpolation becomes unstable: a small change in unit capacitance causes a large change in  \(\sigma\). 
This makes the permeability estimate in the unit capacitance method difficult; thus the penetration depth method will prove to be more reliable at very low porosities.

  A fundamental problem of this study is that of defining the sample boundary. To see
this, note the following paradox: we study the bulk property (permeability) of
a medium using diffusing particles that start outside a finite sample and are absorbed
on its surface. These particles will in general be absorbed in a surface boundary layer that grows thinner as the porosity decreases. Thus, we face the problem of freeing
our simulation from surface artifacts so that it can determine this bulk property, when the method is based on events that occur near the surface! In this paper, we show that this is indeed quite possible.

  We generate samples of the random media models that we study using two different methods: the "effective-radius" method, 
and the "sharp-boundary" method. We describe each method in turn. 

In the "effective-radius" method, if randomly overlapping, i.e., when Poisson statistics are used, we place inclusion sphere centers 
at random in the sample sphere of radius \(R\). 
We continue to place centers until the density of sphere centers approaches \(\rho\), where \(\rho\) and the porosity \(\phi_1\) being modeled are related by:

\begin{equation}
\phi_1=\exp[-(4\pi/3){R^3}\rho]
\end{equation}
If random non-overlapping statistics are used, we place centers sequentially at randomly chosen positions, requiring only that each center placed be far enough from the centers already placed that the corresponding inclusions not overlap. The resulting distribution is not identical to true random non-overlapping statistics, i.e., to what liquid-state theorists call the "hard-sphere" distribution, but studies of these two distributions show that differences in bulk properties emerge only at very low porosities. 
Even though an inclusion extends in part outside the sample sphere 
of radius \(R\), it still contributes 
both to the density of centers \(\rho\) and to the simulation.

  Because of the lack of the contribution
of the inclusion spheres in the region \(r\in[R,R+2a]\), the region \(r\in[R-a,R+a]\) will have
local porosity less than the desired bulk porosity. This is an important concern: especially at lower porosities, the diffusing particles will seldom sample any part of the sample except for the boundary layer. 
Because of this effect, an effective sample radius, which is expected to be less than \(R\), 
is used. For the non-overlapping case, we use for an effective radius the radius of a sphere, which, if it contained the same number of surviving inclusion centers as our sample, would have as its average porosity the same as the bulk porosity of our sample, i.e., the local porosity far away from the sample boundary. (In previous research~\cite{Hwang}, we used this choice of effective radius for both the overlapping and non-overlapping case.) For the overlapping case, the same procedure is used; the set of effective sample radii is then rescaled so that the effective radius at the critical porosity (\( \phi _{1}^{c} \)=0.03) makes the calculated unit capacitance equal to 1. (Here by the term `critical porosity', we mean the porosity below which fluid ceases to flow through the sample, i.e., the percolation threshold.)  

In the "sharp-boundary" method, instead of using an {\it ad hoc} effective radius, the 
porous media sample is constructed as follows: we place the centers of inclusion spheres into a large sphere of radius (\(R + a\)) according to the chosen statistics, for a given porosity. We then define the actual sample by drawing a sphere of radius \(R\) and allowing it to freely intersect inclusions already placed. The sample is then defined to be all of the void phase, all inclusions, and all fragments of inclusions, that are contained in this sample sphere of radius \(R\). 
With this sharp boundary, the porosity in  the actual sample boundary is maintained uniform up to the boundary.

The sampling of periodic grain consolidation models is also done according to the sharp boundary method: we take a spherical sample of radius \(R\)
after making periodic grain models in a cube which is slightly bigger than \(2R \times 2R \times 2R\).
For these models, the result will depend upon the position of the sample center; thus we average the results over ten samples, each with a randomly chosen center.

\section{Studies of Randomly Packed Beds}

  In this section, we apply the two permeability estimation methods to the calculation of 
permeability of model packed beds, composed of non-overlapping, randomly placed,  impenetrable spherical inclusions, 
and model porous media, composed of randomly placed, freely overlapping, impenetrable spherical inclusions. 
We compare our results with the available numerical solutions of the Stokes equation, and also with a number of theoretical bounds and estimates from the literature.

Our simulations show that, for the overlapping sphere model studied here, the two permeability estimation methods, 
the penetration depth method and the unit capacitance method, agree well within random error, estimated, in this case, as the relative standard deviation of the results. 
For the nonoverlapping sphere model, the two methods deviate as porosity decreases. For both models, we obtain 10 packing bed samples and average the ten permeability estimates. 

\begin{figure}
\vspace{0.3cm}
{\centering \includegraphics{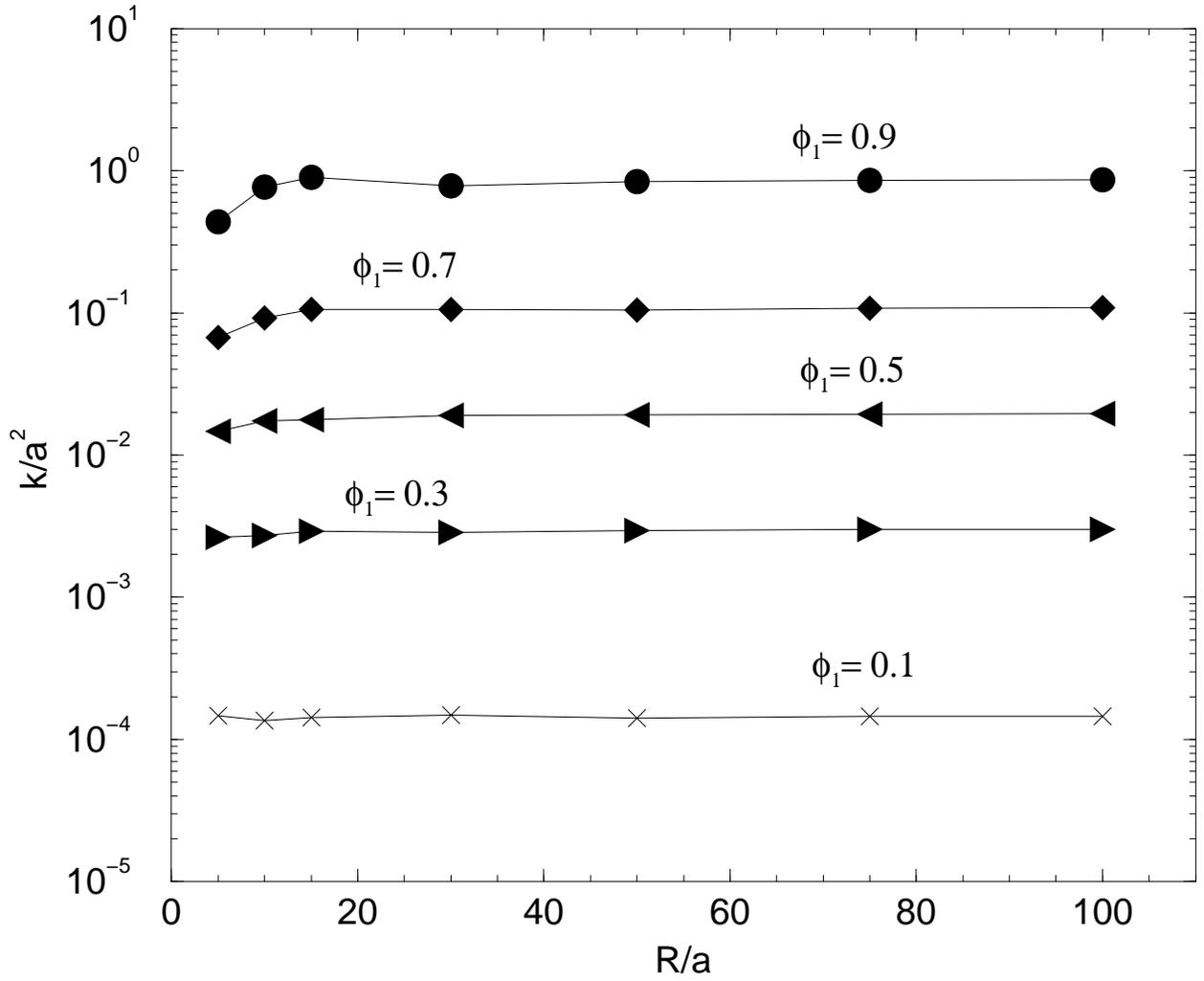} \par}
\vspace{0.3cm}
\caption{ Permeability vs \protect\(R/a\protect\) for penetration depth estimation method applied to the randomly overlapping
sphere model: the permeability estimate does not depend on sample radius, provided that this radius is chosen greater
than a (porosity-dependent) minimum value. The minimum radius increases with porosity.}
\label{Flat1.1}
\end{figure}

As already discussed, the graph of permeability shows a very substantial plateau as a function of sample radius, i.e., a region of sample radius over which the permeability estimate shows almost no variation. 
Also, with the PD method the same property of sampling-size-independent estimation  is observed (see Fig.~\ref{Flat1.1}).

\begin{figure}
\vspace{0.3cm}
{\centering \includegraphics{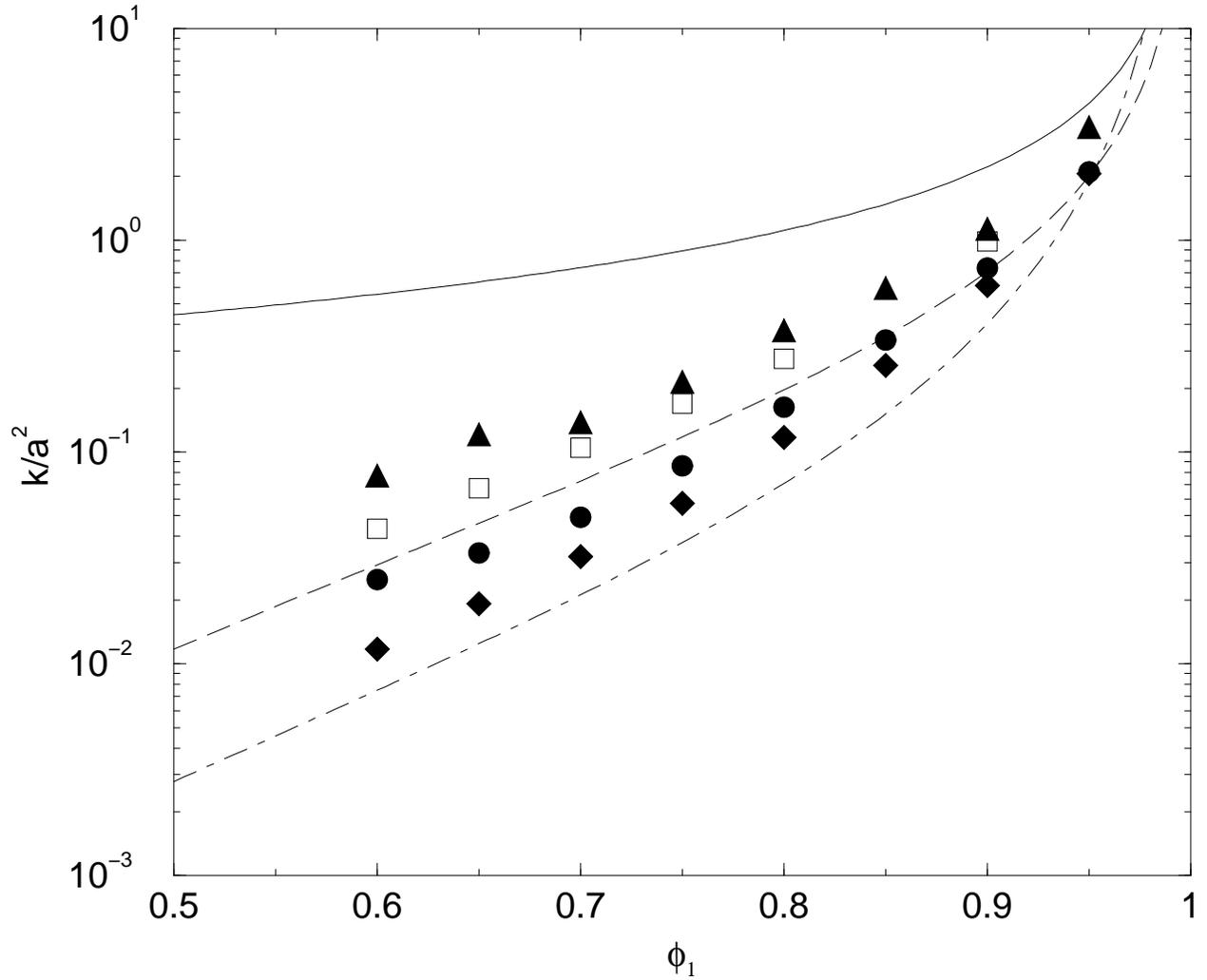} \par}
\vspace{0.3cm}
\caption{Dimensionless permeability \protect \(k/a^{2}\protect\) vs. porosity for the
randomly nonoverlapping sphere
bed (\protect\(a = 1.0\protect\)). The filled upper triangles are our simulation data with the effective radius method,
the filled diamonds are our UC simulation data with the sharp boundary method,
the filled circles are our PD simulation data with the sharp boundary method,
the solid line is the Stokes' law,
the long dashed line is the Happel-Brenner approximation, the squares are the Torquato-Kim
upper bound, and the dot-dashed line is the Kozeny-Carmen relation.}
\label{fig2}
\end{figure}

Our estimation results for the randomly non-overlapping model are compared with some other data sets in Fig.~\ref{fig2}. 
The other data sets used, numbered a) to d),  are:

a) The Stokes' law \cite{HB} for a dilute bed of spheres  is

\begin{equation}
\ k=\frac{2}{9(1-\phi _{1})}a^{2}
\end{equation}

where \( \it a \) is the inclusion sphere radius. 

b) The Happel-Brenner approximation \cite{HB} is,

\begin{equation}
k = (\frac{2}{9 \gamma^{3}}) (\frac{3-\frac{9}{2}\gamma+\frac{9}{2}\gamma^{5}-3\gamma^{6}}{3+2\gamma^{5}})a^2
\end{equation}
where \( \it a \) is the inclusion radius, and \( \gamma^{3}\equiv(1-\phi_{1}) \).

c) The upper bound of Torquato-Kim \cite{Kim-Torquato} is:

\begin{equation}
k = \frac{2}{9 \tau} \frac{\phi_{1}}{(1-\phi_{1})} a^{2}
\end{equation} 

where \( \it D  \) is the diffusion constant and \( \tau \) is the average survival time. The data for \(
\tau \) is taken from Ref. \cite{SB}.

d) The empirical Kozeny-Carmen relation \cite{BK,ST} in the general case is:

\begin{equation}
k = \frac{1}{180} \frac{\phi_{1}^{3}}{(1-\phi_{1})^{2}}a^2
\end{equation}

The Stokes' law, which is the simplest approximation for the dilute bed, gives estimates that are greater than all other data sets.
The Happel-Brenner approximation and the Torquato-Kim upper bound are above our data.
Our simulation data with the effective sample radius is above the Torquato-Kim upper bound.
Our simulation data with the sharp sample boundary lies between the Happel-Brenner approximation and  the empirical Kozeny-Carmen relation.
Because our simulation data with the effective radius method violates the Torquato-Kim upper bound, the sharp boundary method is clearly to be preferred for this case.

\begin{figure}
\vspace{0.3cm}
{\centering \includegraphics{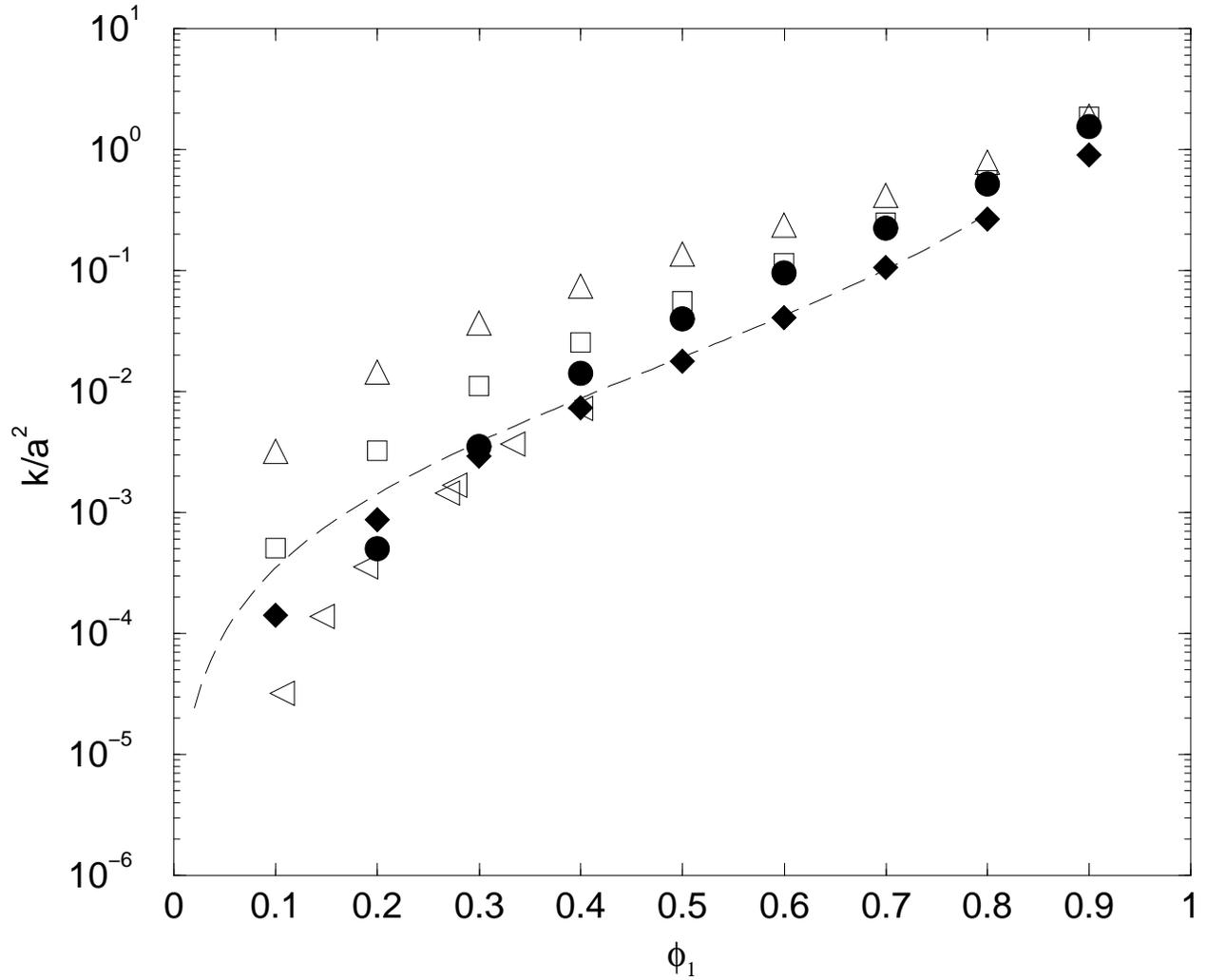} \par}
\vspace{0.3cm}
\caption{The dimensionless permeability k/\protect \( a^{2}\protect\) vs porosity for randomly overlapping spheres
of radius a = 1.0. The filled circles are our simulation data with the effective radius sampling method,
the filled diamonds are our simulation data with the sharp boundary sampling method, the squares are the Torquato-Kim upper bound,
the upper triangles are the Doi upper bound,
the left triangles are the overlapping sphere bed data points from  finite difference solution of the Stokes equation,
and the long dashed line is the lattice-Boltzmann result.}
\label{fig3}
\end{figure}

Our estimation results for the overlapping case \(R = 15.0\) in Fig.~\ref{fig3} are compared with some other data. 
The estimation is not applied at very low porosities (\( \phi_{1}\) <  0.1) 
and very high porosities (\( \phi_{1}\) >  0.9).

The other data sets used, numbered e) to g), are as follows:

e) The Torquato-Kim upper bound on permeability \cite{Kim-Torquato} for an overlapping sphere bed is,

\begin{equation}
k \leq  \frac{D\tau}{F}
\end{equation}

where \(\tau\) is the average survival time and \( \it F \) is the formation factor. 
Data for \( \tau \) and \( \it F \) are obtained from Refs.   \cite{Torquato-Kim} and \cite{Kim-Torquato}.

f) The Doi upper bound~\cite{Doi} on permeability for this system is

\begin{equation}
k = \frac{2}{3}a^{2}e^{-\gamma}\int^{1}_{0}dx(\frac{1}{3\gamma}+x(1-x)^2)\exp(-\gamma(\frac{3}{2}x-\frac{1}{2}x^3)), 
\end{equation}

where \(\gamma\) is given in terms of the porosity by \( \phi_{1} \) = exp(-\( \gamma \)).

g) The lattice-Boltzmann data~\cite{Rothman}, which agrees well with the Kozeny-Carmen equation,

\begin{equation}
\ k = {\phi_{1}}^3/6s^2.
\end{equation}
 
The lattice-Boltzmann method is a variant of the lattice-gas automata method. 
The specific surface area \( \it s \) is given analytically 
by \( \it s = 4\pi a^2 \rho\phi_{1} \) where \( \rho \) is the number of sphere centers per unit volume. At low porosity, finite size effects occur~\cite{MTB}.

Simulation data with the sharp boundary sampling method are below the two upper bounds, the Doi upper bound and the Torquato-Kim upper bound. 
Note that the lattice Boltzmann result at high porosities and the actual data for monosized sphere beds at low porosities agree well with our simulations. 
However, the PD method overestimates
permeabilities at very low porosities because this method does not take into account the phenomenon of percolation.
Our data from the effective radius method is quite good for the overlapping case.

At extremely low porosities, estimation instability appears; when the porosity is very close to unity, 
simulation may give a sample unit capacitance slightly above 1.0 due to the Monte Carlo fluctuations. 
The unit capacitance data which exceed unit value
are discarded. The simulations described here are large enough that this effect is quite small except at some
low porosities of big porous samples like \(R = 50.0\). 

\section{Studies of Periodic Grain Consolidation Models of Porous Media }

 In this section, with the sharp boundary sampling method, we apply the two methods of determining permeability 
that we have developed to 
the permeability calculation of model packed beds, composed of non-overlapping, periodic, impermeable spherical inclusions, 
and model porous media, composed of periodic overlapping, impermeable spherical inclusions. 
The difference between the two types of models consists only in the ratio of the sphere diameter 
to the separation between the spherical inclusions. We compare our results with the solution of the Stokes equation, using fluid dynamics codes, by Larson and Higdon~\cite{Larson}, for simple cubic (SC), body-centered cubic (BCC), and face-centered cubic (FCC) lattices. 

\begin{figure}
\vspace{0.3cm}
{\centering \includegraphics{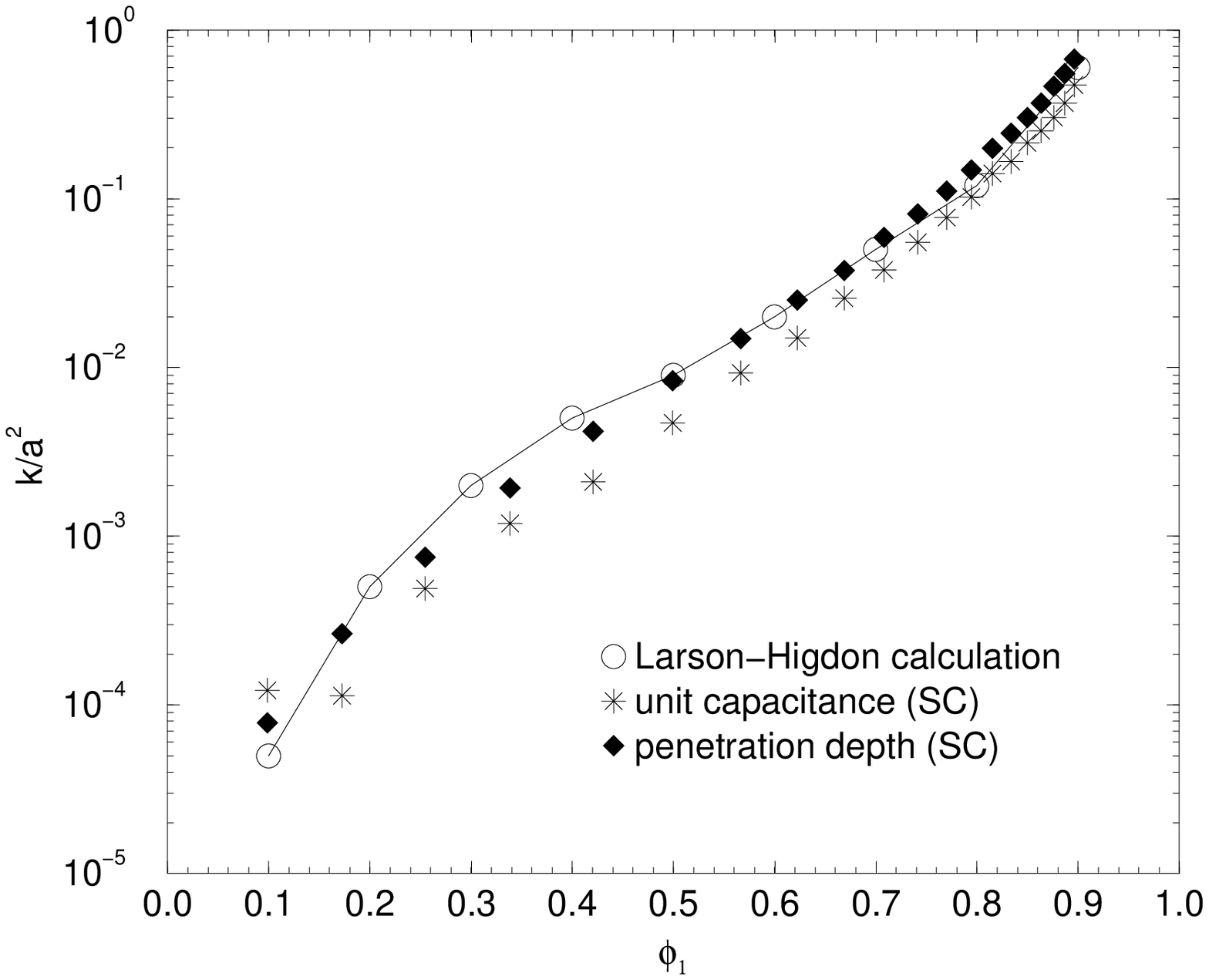} \par}
\vspace{0.3cm}
\caption{The dimensionless permeability \protect  \(k/a^{2}\protect\) vs porosity for SC spheres
of radius \protect \( a = 1.0 \protect \) with \protect \( R = 15.0 \protect \).
The diamonds and stars are our simulation data with the sharp boundary sampling method and the circles are
Larson-Higdon calculation.}
\label{fig0.5}
\end{figure}

\begin{figure}[t]
\vspace{0.3cm}
{\centering \includegraphics{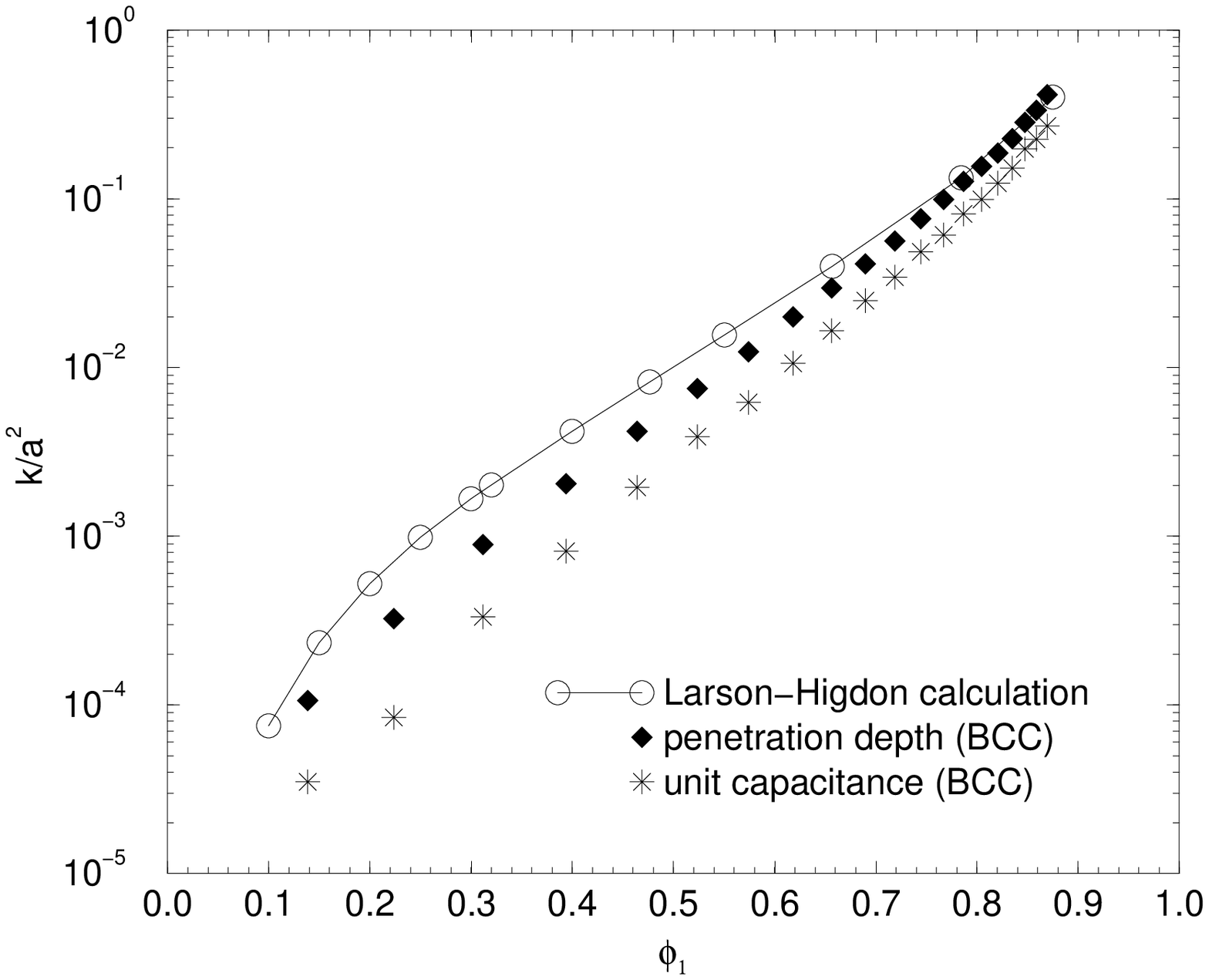} \par}
\vspace{0.3cm}
\caption{The dimensionless permeability \protect \(k/a^{2}\protect\) vs porosity for BCC spheres
of radius \protect \( a = 1.0 \protect \) with \protect \( R = 15.0 \protect \).
 The diamonds and stars  are our simulation data with the sharp boundary sampling method and the circles are
Larson-Higdon calculation.}
\label{fig0.6}
\end{figure}

\begin{figure}[t]
\vspace{0.3cm}
{\centering \includegraphics{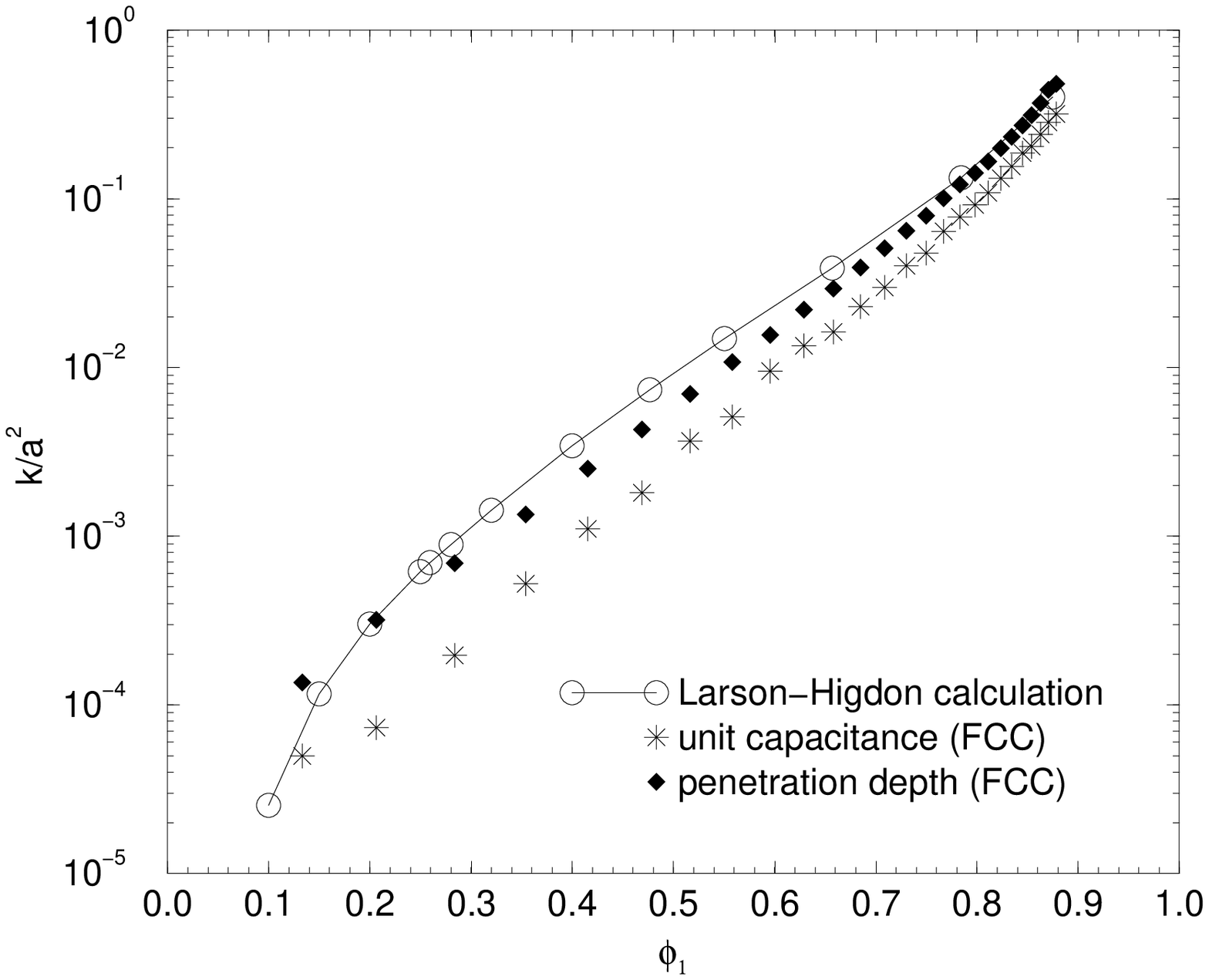} \par}
\vspace{0.3cm}
\caption{The dimensionless permeability \protect \(k/a^{2}\protect\) vs porosity for FCC spheres
of radius \protect \( a = 1.0 \protect \) with \protect \( R = 15.0 \protect \).
The diamonds and stars  are our simulation data with the sharp boundary sampling method and the circles are
Larson-Higdon calculation.}
\label{fig0.7}
\end{figure}

In all cases, 10 packed bed  samples are used, and the results are averaged. Figures~\ref{fig0.5}, ~\ref{fig0.6} and ~\ref{fig0.7} compare results from the two methods with the Larson-Higdon~\cite{Larson} calculations for sample radius \(R=15.0\).
At low porosities, the standard deviation for the unit capacitance estimation is higher because of the estimation instability.
With the  sampling radius \(R = 15.0\), two estimation results, UC and PD, show some deviations. The
PD estimate is better than UC estimate. However, with \(R = 50.0\), the deviation between the two methods becomes smaller. At low porosities, due to the estimation instability of UC, no UC data are shown. It seems that for grain consolidation models UC estimation needs larger sampling due to the cubic lattice structure.

\begin{figure}[t]
{\centering \includegraphics{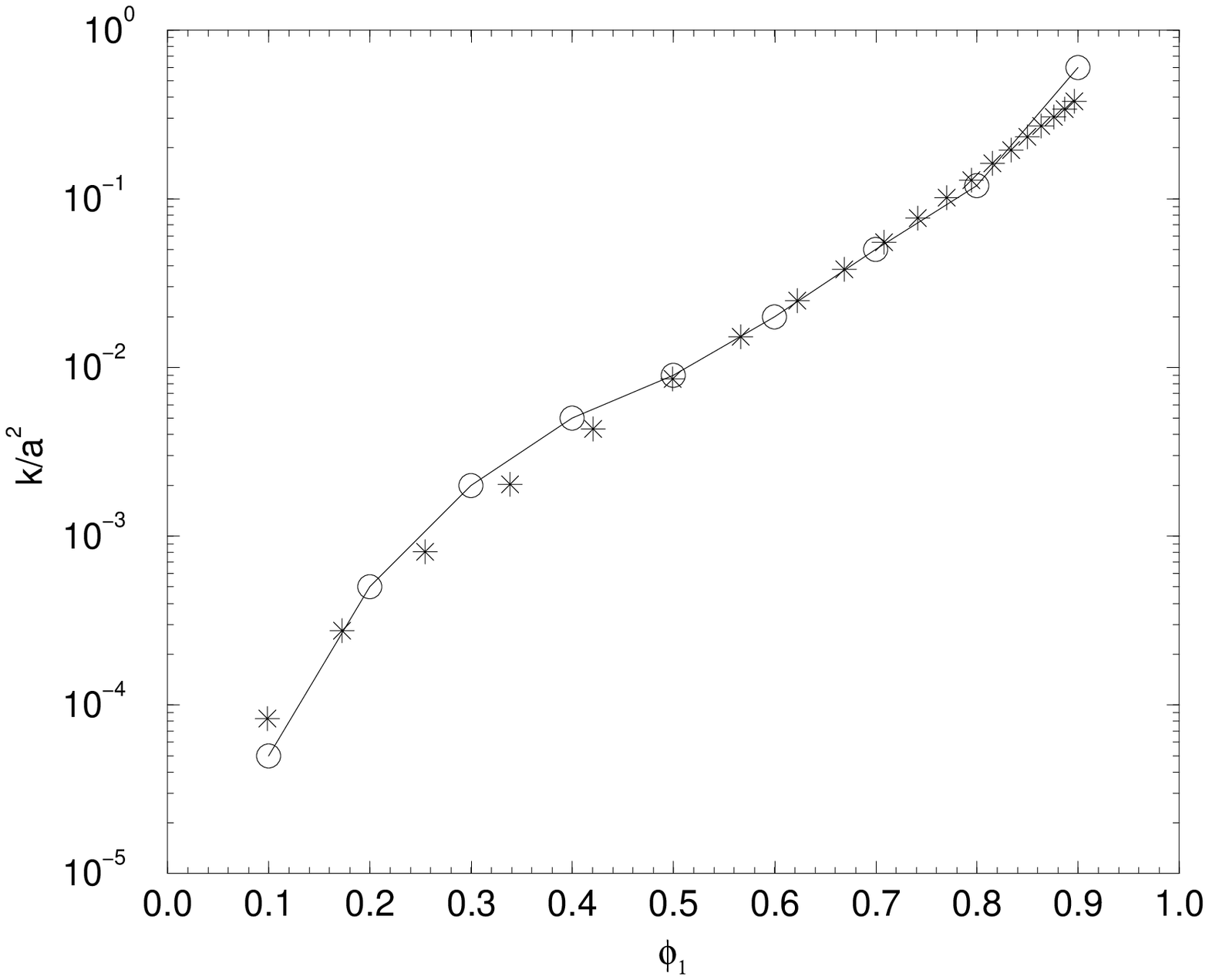} \par}
\caption{Dimensionless permeability \protect \(k/a^{2}\protect\) vs porosity for SC lattice
in the case of \protect \( R = 50.0 \protect \),  \protect \( a = 1.0 \protect \).
The stars are our penetration depth simulation data with the sharp boundary sampling method and the circles are
Larson-Higdon calculations.}
\label{fig4.5}
\end{figure}

\begin{figure}[b]
{\centering \includegraphics{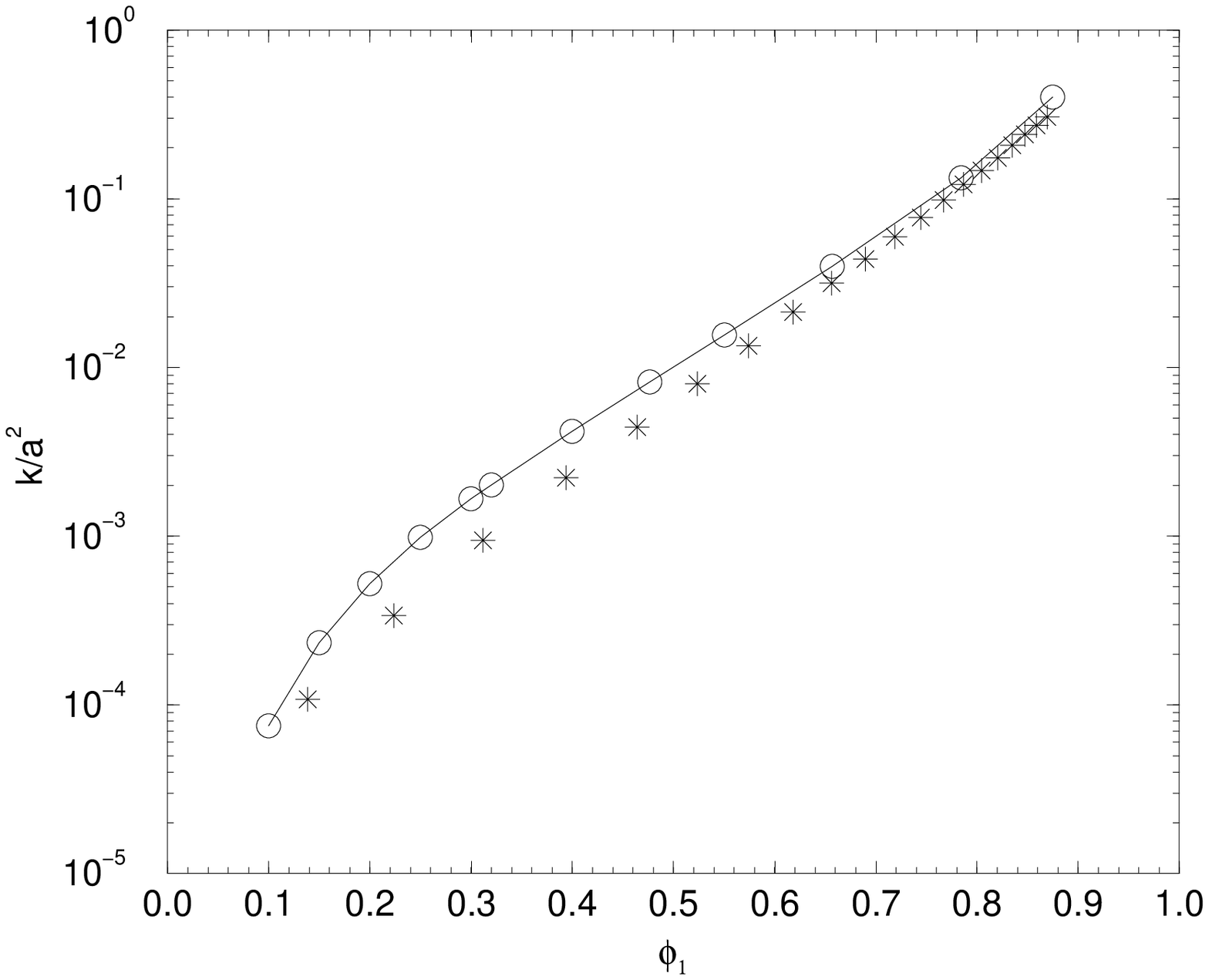} \par}
\caption[Dimensionless permeability \protect \(k/a^{2}\protect\) vs porosity for BCC lattice
in the case of \protect \( R = 50.0 \protect \),  \protect \( a = 1.0 \protect \).]{
Dimensionless permeability \protect \(k/a^{2}\protect\) vs porosity for BCC lattice
in the case of \protect \( R = 50.0 \protect \),  \protect \( a = 1.0 \protect \).
The stars are our penetration depth simulation data with the sharp boundary sampling method and the circles are
Larson-Higdon calculations.}
\label{fig4.6}
\end{figure}

\begin{figure}[t]
{\centering \includegraphics{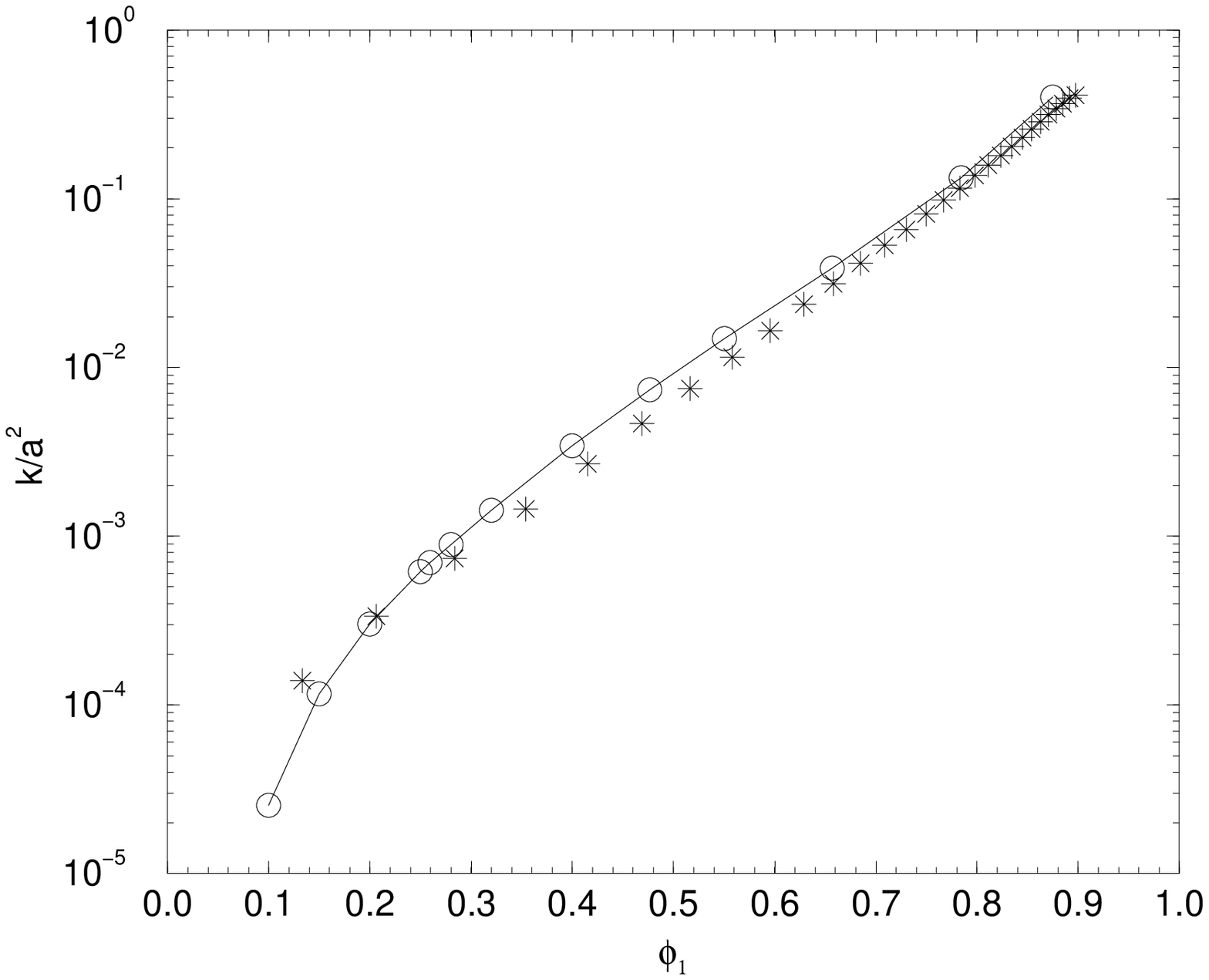} \par}
\caption[Dimensionless permeability \protect \(k/a^{2}\protect\) vs porosity for FCC lattice
in the case of \protect \( R = 50.0 \protect \),  \protect \( a = 1.0 \protect \).]{
Dimensionless permeability \protect \(k/a^{2}\protect\) vs porosity for FCC lattice
in the case of \protect \( R = 50.0 \protect \),  \protect \( a = 1.0 \protect \).
The stars are our penetration depth simulation data with the sharp boundary sampling method and the circles are
Larson-Higdon calculations.}
\label{fig4.7}
\end{figure}

Based on these results, we increase the sampling size to \(R=50.0\). As already discussed, we cannot use the UC method here. The results from the PD method,
Figs.~\ref{fig4.5}, ~\ref{fig4.6}, and~\ref{fig4.7} are in excellent agreement
with the Larson-Higdon results.

\begin{figure}
\vspace{0.3cm}
{\centering \includegraphics{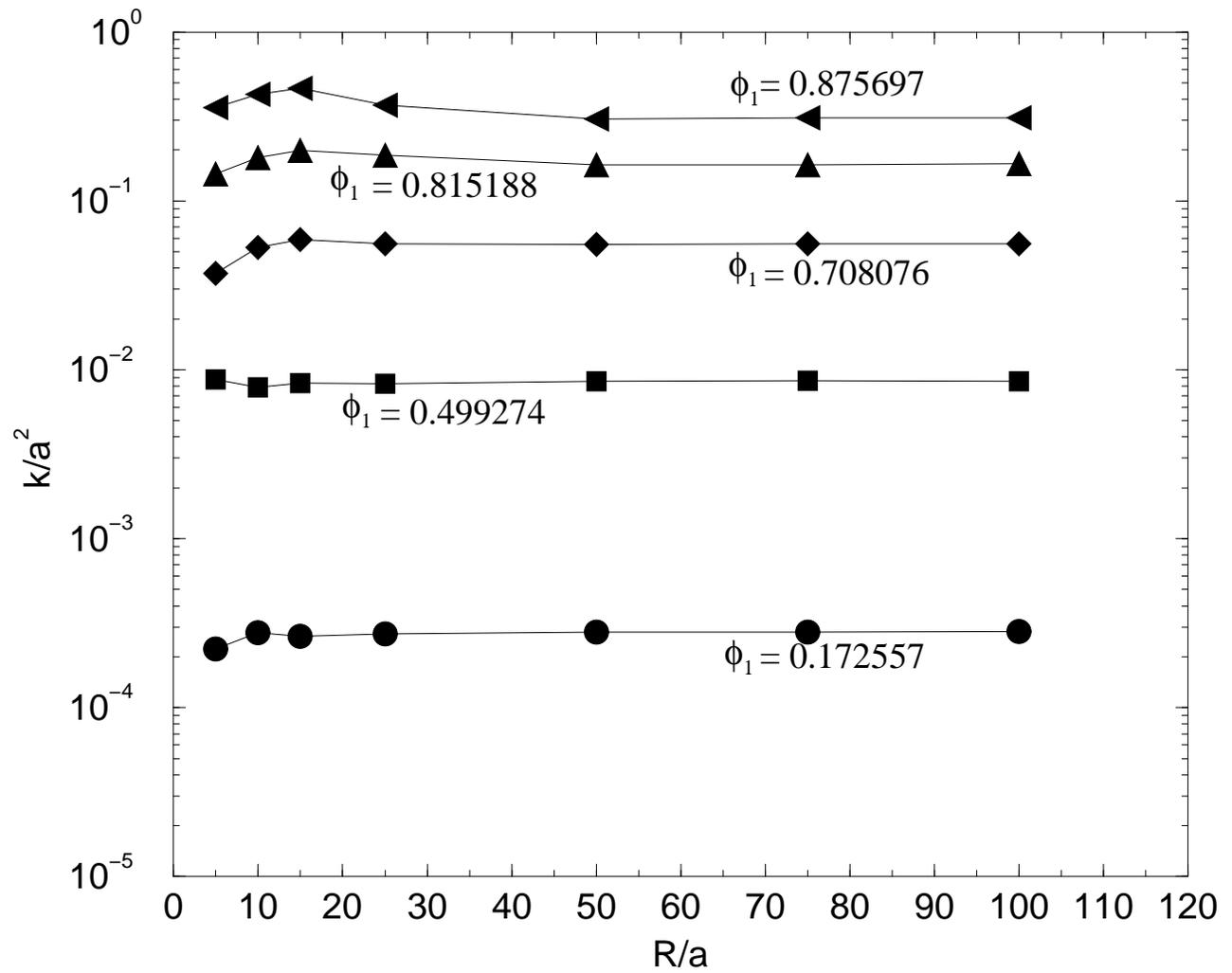} \par}
\vspace{0.3cm}
\caption{Permeability vs \protect\(R/a\protect\) for penetration  depth estimation method
with the sharp boundary sampling method for SC
lattice model. The permeability estimate does not depend on sample radius provided that this radius is chosen greater than a porosity-dependent minimum radius. }
\label{Flat2}
\end{figure}

The same property of sampling-size-independent estimation as the random media  is observed in Fig.~\ref{Flat2}.
It should be noted that, for high porosities, the necessary sample radius can be as large as \(R=50.0\).

\section{Studies of Polydisperse Randomly Porous Media}
  In this section, we apply the two new methods with the sharp boundary sampling method to the calculation of
the permeability of model porous media, composed of polydisperse overlapping, randomly placed, impermeable
 spherical inclusions. The inclusion sphere radii are chosen at random from the values \{1.5, 3.5, 5.5, 7.5\}. 
We compare our results with the available numerical solutions of the Stokes equation~\cite{MTB}.

\begin{figure}[b]
\vspace{0.3cm}
{\centering \includegraphics{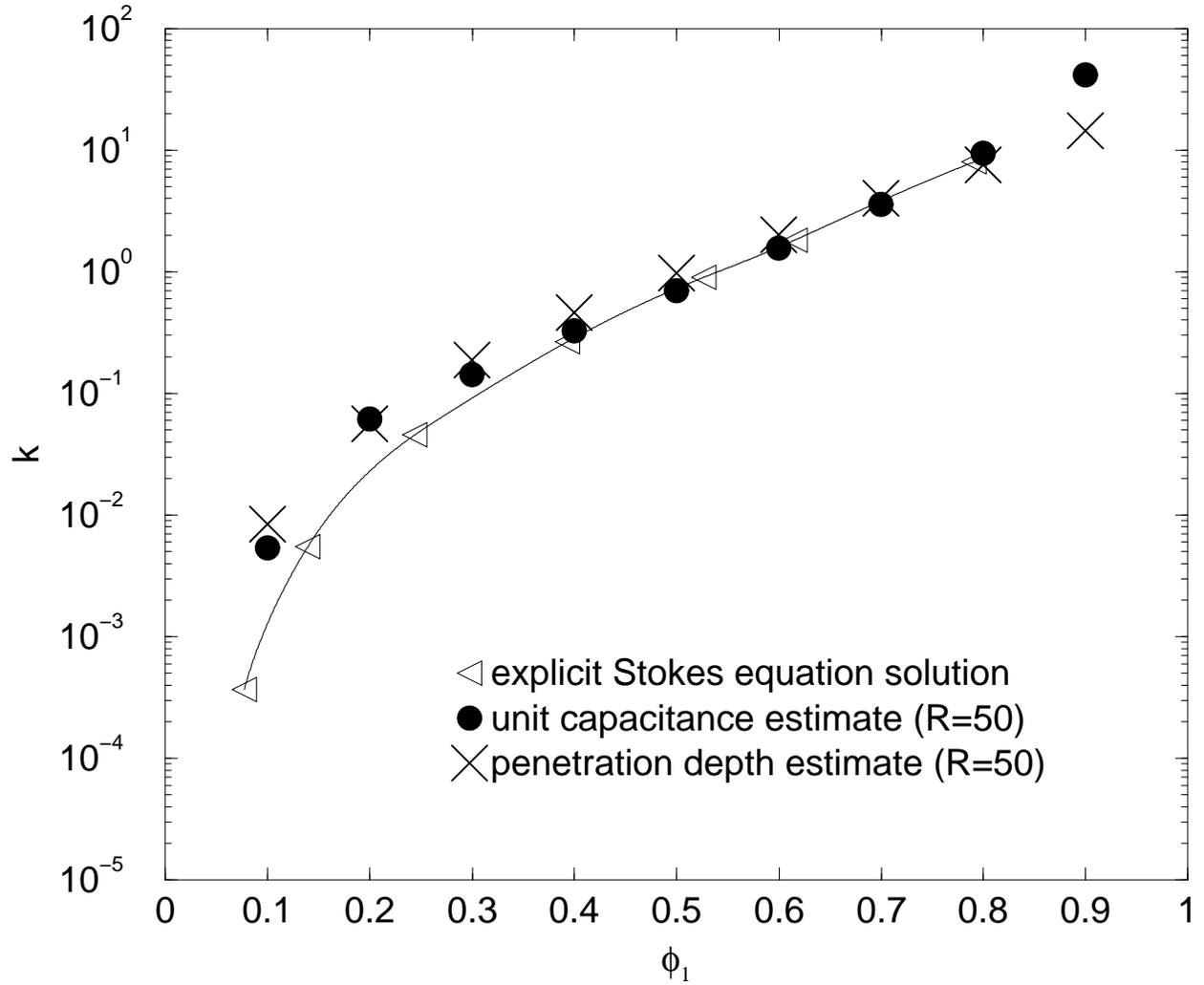} \par}
\vspace{0.3cm}
\caption{Permeability \protect \(k\protect\) vs porosity for a matrix constituted by a polydisperse mixture of randomly overlapping impermeable spheres with the sharp boundary sampling method; the sphere radii are chosen to
have the four values \protect \( a = \{1.5, 3.5, 5.5, 7.5\}
\protect \) with equal probability. Here the sample radius
\protect \( R = 50.0 \protect \).}
\label{Poly_Over}
\end{figure}

Figure~\ref{Poly_Over} shows that the two sets of results agree well. At very low porosities, our methods give permeabilities greater than those given by the Stokes equation because we have not yet incorporated the phenomenon of percolation. This will become important when we perform
more extensive study of polydisperse models; the percolation threshold seem to be reached at lower porosities in such systems. 

\section{Conclusions and Suggestions for Further Study}

Our methods give very good results for all models of porous media tested. 
The penetration length method is better at low porosities; the unit capacitance method
shows high standard deviations at low porosities due to the steepness of the unit capacitance curve.
It is very unfortunate that so little high quality simulation data exists even for the simple models studied here. 
We note that these models have been standards for theoretical study for decades. Our method 
will predict permeabilities for a large class of homogeneous and isotropic porous media, in the medium and 
high porosity regimes \((0.1 < \phi_{1} < 0.9)\). 

The unit capacitance estimation discrepancy with sampling radius \(R=15.0\) may be due to the importance, 
for periodic models, 
of using a launch surface with the geometry of the sample, e.g., a cubic launch surface for SC samples. 
We have laid the basis for this step here; it will require further study.  

  An important point of this method is that it is very fast compared with other methods. 
Our computations were in parallel Fortran, using MPI, on a PC cluster (32 node, 233 Mhz machine).
Using 10 nodes, one porous sample per node, one million diffusing particles for each porous sample and porosity, 
and 10 porosities, it takes about 10 hours per set of calculations. 
Using e.g. boundary element/finite element methods to solve the Stokes equation in a sample of porous media can 
require the same amount of time to do this set of calculations for a single value of porosity 
(We note that the latter methods are, at present, more efficient than these methods developed 
here for obtaining the detailed flow field in a sample; this may be important in particular 
applications.).

 {\centering \textbf{ACKNOWLEDGMENTS} \par}
We give special thanks to Nicos Martys, Eduardo Glandt and Dan Rothman for their useful information and comments. We thank Joe Hubbard and Jack Douglas for useful discussions. We are especially grateful to Nicos Martys for sharing with us the raw simulation data used in reference~\cite{MTB}.

Also, we wish to acknowledge NCSA, especially the Condensed Matter Physics section of the Applications Group, for access to their computational resources and collaborative support.  In addition, we thank NCSA's educational programs for training support to use NCSA computational resources.

\end{document}